# Non-linear Landau fan diagram for graphene electrons exposed to a moiré potential

Pilkyung Moon[1,2, △,*], Youngwook Kim[3,4, △], Mikito Koshino[5,△,*], Takashi Taniguchi[6], Kenji Watanabe[7], and Jurgen H. Smet[3,*]

[1]*Arts and Sciences, NYU Shanghai, Shanghai 200124, China*

[2]*NYU-ECNU Institute of Physics at NYU Shanghai, Shanghai 200062, China*

[3]*Max-Planck-Institut für Festköperforschung, 70569 Stuttgart, Germany*

[4]*Departement of Physics and Chemistry, DGIST, 42988, Korea*

[5]*Department of Physics, Osaka University, Toyonaka 560-0043, Japan*

[6]*International Center for Materials Nanoarchitectonics, National Institute for Materials Science, Tsukuba 305-0044, Japan*

[7]*Research Center for Functional Materials, National Institute for Materials Science, Tsukuba 305-0044, Japan*

△These authors contributed equally

*E-mail: pilkyung.moon@nyu.edu, koshino@phys.sci.osaka-u.ac.jp, and j.smet@fkf.mpg.de

**Abstract:**

**Due to Landau quantization, the conductance of two-dimensional electrons exposed to a perpendicular magnetic field exhibits oscillations that generate a fan of linear trajectories when plotted in the parameter space spanned by density and magnetic field. This fan looks identical irrespective of the electron dispersion details that determines the field dependence of the Landau level energy. This is no surprise, since the position of conductance minima solely depends on the level degeneracy which is linear in flux. The fractal energy spectrum that emerges within each Landau band when electrons are also exposed to a two-dimensional superlattice potential produces numerous additional oscillations, but they too create just linear fans for the same reason. Here, we report on conductance oscillations of graphene electrons exposed to a moiré potential that defy this general rule of flux linearity and attribute the anomalous behavior to the simultaneous occupation of multiple minibands and magnetic breakdown.**

The spectral gaps in the density of states that originate from Landau quantization of the energy spectrum of a two-dimensional electron system exposed to a perpendicular magnetic field produce a fan of linear trajectories in a contour plot of the transport quantities in the parameter plane spanned by density and field [1]. The linear trajectories are described by a simple Diophantine equation of the form $n/n_0 = t\, \phi/\phi_0 + s$. Here, $\phi/\phi_0$ is the normalized magnetic flux and $n/n_0$ is the density normalized to the total density $n_0$ that can be accommodated by the partially filled band, ignoring any extra degeneracies associated with for instance spin and valley degrees of freedom. The integers $s$ and $t$ are topological integers representing the band filled at zero field and the quantized Hall conductivity, respectively [2-5]. This outcome is independent of the size and the field dependence of the gaps separating adjacent Landau levels, since only the level degeneracy matters and it just depends on the number of magnetic flux quanta that pierce the 2D system. Hence, it is not possible to extract any information about the energy spectrum such as the energy spacing of Landau levels from such a density-field or Wannier diagram. To obtain the gap sizes, we need to perform either cyclotron resonance studies or thermal activation experiments, which measure how much thermal energy is required to overcome the gap closest to the chemical potential.

When 2D electrons are subjected to a 2D superlattice potential, Bragg scattering and zone folding converts the original bands in a series of much smaller minibands [6]. Each miniband of the superlattice can host a density $n_0 = 1/A$, where $A$ is the area of the unit cell, or an integer multiple thereof, if carriers have additional degrees of freedom. This density $n_0$ can be much smaller than the density a conventional crystal band holds. Hence, multiple minibands can easily be filled through electrostatic gating. Additional linear Landau fans emanate on the density axis whenever a miniband is completely filled or emptied. In addition, the superlattice potential actually broadens these Landau levels into bands that themselves develop $p$ internal subbands of equal weight, whenever the flux penetrating a unit cell of the periodic potential takes on a rational multiple $p/q$ of the flux quantum $\Phi_0 = h/e$. Here, $p$ and $q$ are relative primes. This results in a fractal spectrum within each Landau band referred to as Hofstadter's butterfly [2-5,7-12]. The additional gaps separating these subbands also generate linear trajectories in the density-field plane and are again described by a Diophantine equation [1]. As before, it is not possible to extract the energy spacing of Landau levels, since the conductance minima appear at a density and field determined by the degeneracy of the electronic states and this position has no bearing on the size of the gap.

The successful isolation of single atomic sheets from layered materials and in particular the ability to combine such sheets through van der Waals stacking into a heterostucture offers unprecedented control to create such devices where electrons are exposed to a superlattice potential with a periodicity of a few to a few tens of nanometer [2-5,7-14]. The superlattice potential is caused by the moiré interference potential that emerges when two identical atomic layers are stacked at an angle or when two distinct layers with some lattice mismatch are placed on top of each other. Here, we have investigated graphene electrons that are subjected to the moiré potential created by a hBN layer aligned with the graphene lattice. While the magnetotransport features in a Wannier diagram are indeed dominated by linear trajectories formed by conductance minima due to Landau quantization of the minibands as well as the internal fractal Landau band structure, also unanticipated non-linear trajectories are observed that do not fulfill a flux linear Diophantine equation. This anomalous behavior appears in the density regime where more than just one miniband is partially occupied and the Fermi surface areas, that these partially filled minibands contribute, are very different. We demonstrate that these non-linear trajectories offer the unique opportunity to extract information about the spectral gaps. This technique to obtain energy information should be applicable to a broader class of systems such as multilayer graphene with various stacking configurations (AA, AB, ABC, …) as well as twisted bilayer graphene.

The magnetoresistance data reported here were recorded on two hBN-encapsulated monolayer graphene devices, referred to as D1 and D2. In both, the top hBN layer was approximately aligned with the graphene in order to generate a moiré potential with a period of about 13.9 nm (Section 2 of the Supporting Information). The second hBN layer at the bottom was intentionally misaligned in order to avoid another moiré superlattice. The doped silicon substrate served as the back-gate in device D1, while a graphite layer was used as top and bottom gate in device D2. Since the transport behavior is consistent among both devices, we only show results from D1 in the main text. The data acquired on D2 is deferred to Section 3 of the Supporting Information (SI). Details of the device fabrication can be found in the Method Section. The bottom panels of Fig. 1a and b display the longitudinal resistance $R_{xx}$ at zero field ($B = 0$) and the Hall resistance at $B = 0.2$ T as a function of the density normalized to the total density that can be hosted by a single superlattice miniband, $n_0$, when ignoring other degeneracies. Also shown are the electronic band structure (Fig. 1c) along high symmetry points of the superlattice Brillouin zone (Fig. S3) and density of states (Fig. 1d) calculated using the effective continuum model for a graphene/hBN heterostructure with a 0° twist [15]. As anticipated for electrons exposed to a moiré potential, the longitudinal resistance exhibits three peaks. The maxima at $n = \pm 4n_0$, signal full occupation or depletion of the miniband centered around zero energy. Here, the factor of 4 accounts for the spin and valley degrees of freedom. The resistance peak at $n = 0$ corresponds to the charge neutrality point and is a result of the vanishing density of states as the chemical potential approaches the Dirac point. The sign reversal of the Hall resistance at these three characteristic densities is consistent with this interpretation. Since Bragg scattering of the superlattice produces a van Hove singularity on either side of the miniband (Figs. 1c and 1d), an additional sign reversal of the Hall resistance occurs due to a Lifshitz transition, i.e. an abrupt change of the Fermi surface topology, when the chemical potential crosses these singularities [16, 17]. The excellent agreement between these experiments and the theoretical calculations unequivocally proves that electrons are subjected to the moiré superlattice induced by the hBN aligned and in close proximity with graphene.

The top graphs in Fig. 1a and 1b show color maps of the longitudinal conductivity $\sigma_{xx}$ and Hall conductivity $\sigma_{xy}$ as a function of back-gate voltage $V_g$ or density and perpendicular field $B$. The conductivities are obtained from the inversion of the resistance tensor, $\sigma_{xx} = R_{xx}(w/L) / ((R_{xx}(w/L))^2 + R_{xy}^2) \times R_K$ and $\sigma_{xy} = R_{xy} / ((R_{xx}(w/L))^2 + R_{xy}^2) \times R_K$. Here, $w$ and $L$ are the sample width and length, respectively. $R_K$ is the von Klitzing constant ($\approx 25{,}812$ Ω). Minima in $\sigma_{xx}$, which track the spectral gaps, are visible down to 0.8 T. They appear as straight lines in this Wannier diagram and follow the Diophantine equation. They converge to the main charge neutrality point at zero energy (main CNP) or to the top or bottom of the miniband, referred to as mini CNPs as $B$ approaches zero. A portion of the experimental data in Fig. 1a is replotted in Fig. 2a with an optimized color scale near $n/n_0 = 4$ ($V_g \approx 40$ V) in order to bring this out better. A color map of the conductivity that extends to higher fields is included in Section 1 (SI). It reveals the Hofstadter butterfly spectrum [7].

All of the above discussed features in the Wannier diagram have been addressed previously in the literature [2-5]. In this work, we focus on an unusual sequence of minima in $\sigma_{xx}$ that appears at weak fields ($B < 3$ T) when more than just the lowest miniband is occupied ($n/n_0 > 5$ or $V_g > 50$ V). As seen in Fig. 1a and even more so in Fig. 2a and 2b (40 V $< V_g <$ 80 V), the minima in this regime trace trajectories that are distinct from the common linear Landau fan. The trajectories are not linear in the ($n,B$)-plane, but parabolic-like and they do not converge to one of the nearby band edges at $n/n_0 = 4, 8$ ($V_g = 40$ V, 80 V) when the field approaches zero. Note the overall high value of the conductivity (Fig. 2a), exceeding 100 $e^2/h$, even at the minima. This is much higher than for the previously discussed minima associated with either Landau quantization or the moiré lattice induced internal Landau band gaps. These non-linear features weaken as the field is raised and they disappear near 3 T. The SI contains a data set for D2 in Section S3 and similar non-linear trajectories appear for the same density range. Since within a relaxation time approximation the conductivity is approximately proportional to the density of states, it should come as no surprise that a calculation of the latter for this aligned graphene/hBN superlattice shows strong similarities to the conductivity map in Fig. 2a. Fig. 2c and 2d display a color rendering of the density of states as a function of $B$ and as a function of energy $E$ (panel c) or normalized density (d). For simplicity, a Gaussian broadening with a constant full width at half maximum of 0.5 meV was applied. The resemblance of the experimental data for densities $n/n_0 > 5$ is striking. Minima in the density of states trace non-linear trajectories in the ($n,B$)-plane (Fig. 2d) and these do not converge to $n/n_0 = $ 4 or 8 as the field approaches zero, consistent with the experimental data.

In order to identify the origin of the non-linear features in the conductivity minima and the density of states map, it is instructive to study the isoenergetic energy contours at different Fermi energies for this graphene/hBN superlattice. The top panel in Fig. 3a shows the dispersion of the first three conduction minibands near the *K*-symmetry point for $B = 0$ T calculated with the effective continuum model [15]. These minibands are not separated in energy. While the first miniband (green) only slightly overlaps with the second (orange), the second miniband overlaps with the third miniband over an extended range of energies when $E > 0.15$ eV. This is also apparent from the dispersions of these minibands along the high symmetry points in Fig. 1c, where these minibands were color-coded in the same fashion as in Fig. 3a. The isoenergetic contours are plotted for six different Fermi energies in Fig. 3b. For $E > 0.15$ eV, the third miniband develops pockets centered around the *X* and *Y* symmetry points of the reduced Brillouin zone. The contours encircling an *X*-symmetry point (purple shaded lines in the panel with $E = 0.18$ eV) are well separated in reciprocal space from the isoenergetic contours of the second miniband. The quantization of the corresponding real space orbit in a magnetic field should therefore produce a set of discrete Landau levels. In the remainder, we will refer to these obits as *α*-orbits (purple shaded lines in Fig. 3b). The contours encompassing a *Y*-symmetry point on the other hand nearly touch the isoenergetic contours of the second miniband. Under these circumstances, electrons have a finite probability of tunneling between this orbit of the third miniband and the orbit associated with the second miniband (yellow shaded lines in the panel with $E = 0.18$ eV). In general, as the field increases, real space orbits shrink. Hence, the uncertainty in real space drops at the expense of an increased uncertainty in momentum space. The latter scales proportional to $\sqrt{B}$ and enables the tunneling. This effect, well-known as magnetic breakdown [18], effectively converts both orbits involved into open snake orbits, that we hereafter refer to as *β*-orbits (along yellow shaded lines in Fig. 3b, one snake orbit running in the vertical direction is highlighted with a dotted line and arrows). These open *β*-orbits do not contribute well separated discrete Landau levels, but produce a nearly continuous contribution to the spectrum and the density of states. Figures 4e and f summarize schematically how the energy spectrum and density of states looks like in this regime where magnetic breakdown is active. The energy spectrum is composed of the well separated Landau levels in purple with energies $E_i^\alpha$ due to the Landau quantization of the real space *α*-orbits as well as narrowly spaced levels marked in yellow that stem from the *β*-orbits. In the density of states, all Landau levels are broadened with a Gaussian so that the yellow levels make up a nearly continuous background. The *α*- and *β*-orbits virtually don't mix even at higher energies as apparent from the simple crossing of these energy contours encircling the *X*- and *Y*-symmetry points at $E =$ 0.23 eV; they persist and do not interfere almost up to $n/n_0 = 12$, i.e. the complete filling of the third miniband.

The co-existence of a nearly continuous spectrum due to the $\beta$-orbits with a spectrum of well separated Landau levels due to the $\alpha$-orbits is at the core of the observed non-linear features in the Wannier diagram. To corroborate this assertion we derive the Wannier diagram for the above described spectrum and compare it with a system that has an energy spectrum just composed of a single series of Landau levels (Fig. 4a and b). In general, the Landau levels are not necessarily equidistant. The electron density to completely fill each Landau level in Fig. 4b depends solely on the degeneracy, which, irrespective of any details of the band structure, is equal to the number of flux quanta that thread the sample (Fig. 4c). Therefore, although these Landau levels are not equally spaced in energy, they appear equidistant in the $(n,B)$-plane (Diophantine equation, Fig. 4d) and produce linear trajectories of the conductivity minima in the Wannier diagram. For the energy spectrum shown in Fig. 4e, on the other hand, the change in density required to completely fill the next level associated with orbit $\alpha$ is no longer solely determined by its degeneracy, since additional states of the nearly continuous spectrum from $\beta$-orbits also need to be filled. For instance, the required change in the total density to raise the chemical potential from the peak in the density of states at $E_i^\alpha$ to the neighboring peak at $E_{i+1}^\alpha$ is

$$\Delta n \equiv n(E_{i+1}^\alpha) - n(E_i^\alpha) = \int_{E_i^\alpha}^{E_{i+1}^\alpha} \left( D_\alpha(E) + D_\beta(E) \right) dE = g n_L + \int_{E_i^\alpha}^{E_{i+1}^\alpha} D_\beta(E) \, dE, \qquad (1)$$

where $n(E)$ is the total density at energy $E$, $D_\alpha$ and $D_\beta$ represent the density of states for the $\alpha$- and $\beta$-orbits, respectively, $n_L = eB/h$ is the degeneracy of a single Landau level and $g$ is the additional degeneracy due to degrees of freedom other than the orbital one. When $D_\beta(E)$ is zero, the spectrum is composed only of a single series of Landau levels; Eq. (1) then just yields $g n_L$ and full Landau level occupation generates the usual linear Landau fan in the $n$-$B$ plane, independent of the central energies $E_i^\alpha$ as shown in Fig. 4d. For non-zero $D_\beta(E)$, however, the last term in Eq. 1 becomes relevant and will in general cause non-linear trajectories as seen in Fig. 4h. We note that panels d and h highlight conductivity maxima, since $E_i^\alpha$ and $E_{i+1}^\alpha$ refer to adjacent maxima in the density of states. In principle we can use Eq. (1) also for conductivity minima by changing the integration limits to energies in between the $\alpha$-levels. From now on, we will follow the conductivity maxima in order to directly trace $E_i^\alpha$. Since the second term in Eq. 1 scales with the energy spacing between adjacent levels, we can reverse engineer the level spacing from the Wannier diagram, provided $D_\beta$ is known or can be estimated. In other words, we can use $D_\beta$ as a measure of the energy spacing $\Delta E_i^\alpha = E_{i+1}^\alpha - E_i^\alpha$. Such reverse engineering is not possible in a system with a single series of discrete Landau levels, which has zero $D_\beta(E)$. If $D_\beta(E)$ slowly varies with energy, we can get the level spacing from

$$\Delta E_i^\alpha \approx (\Delta n - g n_L)/D_\beta\left(\frac{E_{i+1}^\alpha}{2} + \frac{E_i^\alpha}{2}\right), \qquad (2)$$

a formula that indeed only works for $D_\beta \neq 0$. In principle, we can use Eq. 2 to probe the Landau level spectrum of any 2D material proximitized with a reference material with much smaller level spacing and a known density of states (e.g., metal). Or, we can extract the Landau level spectrum in a system having two different series of Landau levels in the same energy range but with very different level spacing.

With this background and Eq. (2) it is possible to extract the Landau level spectrum generated by the $\alpha$-orbits from the data. While $D_\beta(E)$ is unknown in our case, we can assume that the total density of states $D_\alpha + D_\beta$ at any $E$ is close to that of an intrinsic graphene monolayer, since the level spacing is small at moderate $B$. The total density $n$ at Fermi energy $E$ is then simply given by $n(E) = gE^2/(4\pi\hbar^2 v^2)$, where $v$ is graphene's band velocity and $g = 4$ is the valley and spin degeneracy. By defining the inverse function $E(n) = \hbar v\sqrt{4\pi n/g}$, we can convert the trajectory of a conductivity maximum in the Wannier diagram to the energy of a Landau level $E^\alpha(n(B))$. Fig. 5 displays the Landau level spectrum extracted from the data in Fig. 1a in this manner. Note that we can only extract the energy spacing between adjacent Landau level states and not their absolute energy value, since we do not know the energy of the first, i.e. lowest level.

In conclusion, we have identified conductivity oscillations that trace anomalous non-linear trajectories in a Wannier diagram and do not converge to any miniband edge in an aligned graphene/hBN heterostructure. They appear when multiple Fermi surfaces co-exist and when magnetic breakdown creates real space orbits with very different areas. These non-linear trajectories can be exploited to extract the magnitude and field dependence of spectral gaps in the energy spectrum of the two dimensional electron system from the Wannier diagram, normally an impossible feat.

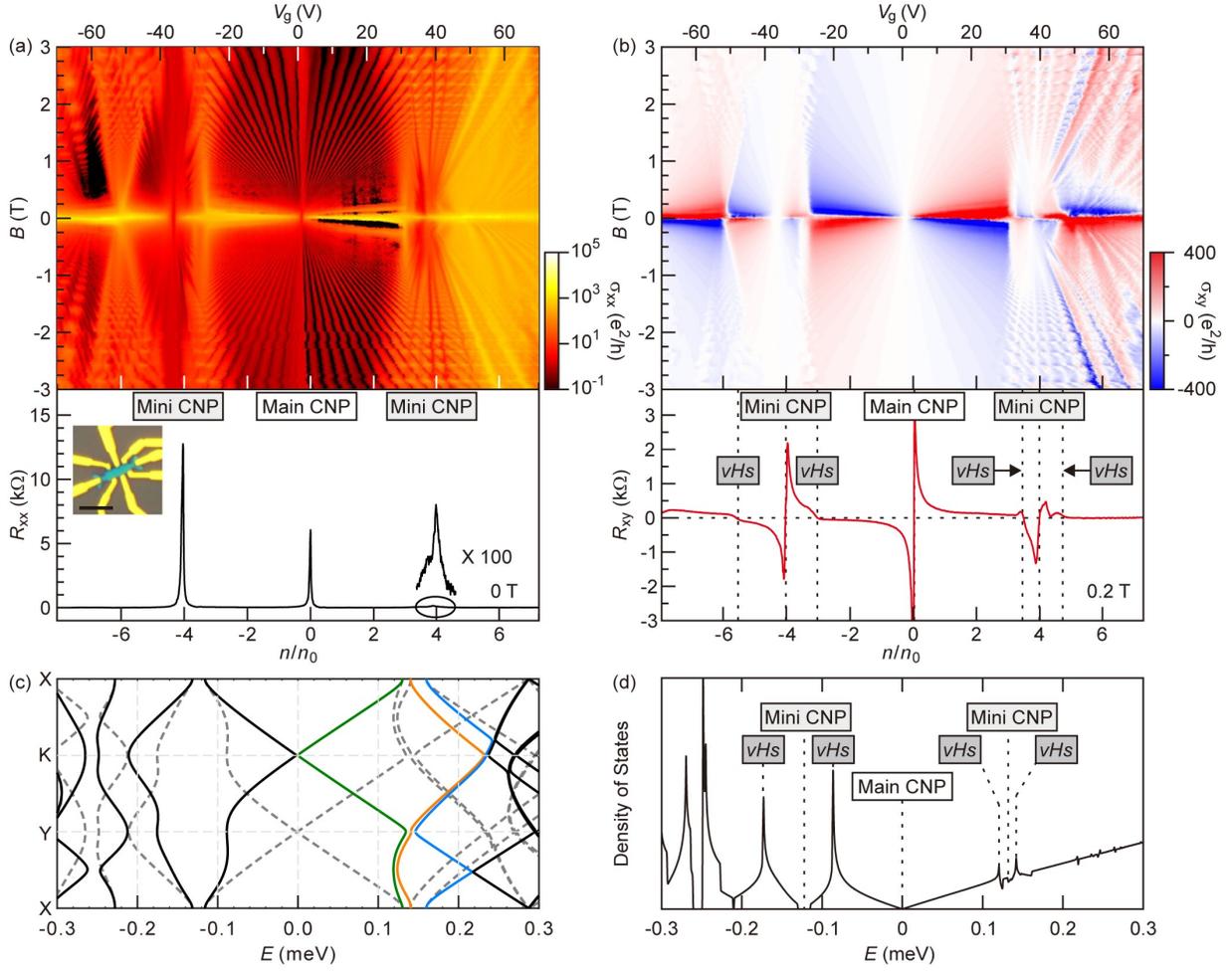

Figure 1. Magnetotransport of a graphene/hBN heterostructure with a 0º twist angle. (a) Top: $\sigma_{xx}$ as a function of gate voltage ($V_g$) or normalized density ($n/n_0$) and magnetic field ($B$) for a logarithmic color scale and fields up to ±3 T. $V_g$ and $B$ are increase in 5 mV and 10 mT steps. The data are recorded at about 30 mK. Bottom: $R_{xx}$ as a function of $n/n_0$ for $B$ = 0T. The inset shows an image of the device. The scale bar corresponds to 5 μm. (b) Top: Similar as (a) but for $\sigma_{xy} = R_{xy} / ((R_{xx} (w/L))^2 + R_{xy}^2) \times R_K$. Bottom: Single line trace of $R_{xy}$ at $B$ = 0.2 T. The horizontal dotted line corresponds to $R_{xy}$ = 0 Ω and the vertical dotted lines mark zero crossing or charge inversion points. Charge inversion occurs when the chemical potential reaches the Dirac point (main CNP), a band edge (mini CNP) or a van Hove singularity (vHs). (c) Band dispersions along the high symmetry points of Fig. S3 for the graphene/hBN heterostructure with zero twist angle. Solid (dotted) lines represent the bands near the $K$ ($K'$) valley of monolayer graphene. Green, orange, and blue highlight the dispersion of the first, second, and third conduction minibands, respectively. (d) Density of states plotted against the electron energy.

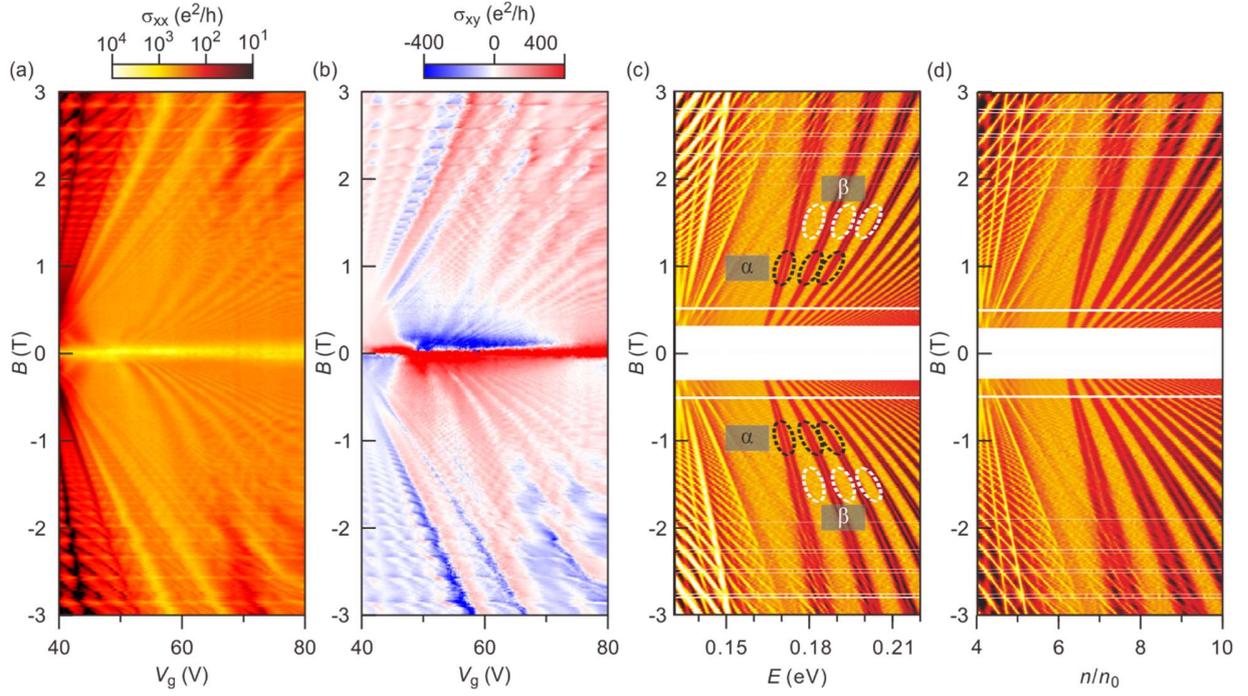

Figure 2. Comparison between the magnetotransport properties and the density of states of the graphene/hBN heterostructure with zero twist angle. (a) $\sigma_{xx}$ measured as a function of $B$ and $V_g$. (b) Same as (a) but for $\sigma_{xy}$. (c) Theoretically calculated density of states in the $E$ versus $B$ plane. The black and white oval shaped dotted lines mark levels due to Landau quantization of the $\alpha$-orbits of the third miniband and the snake or $\beta$-orbits that appear as a result of magnetic breakdown among closed Fermi contours of the second and third miniband. (d) Same as (c) but in the $n/n_0$ versus $B$-plane.

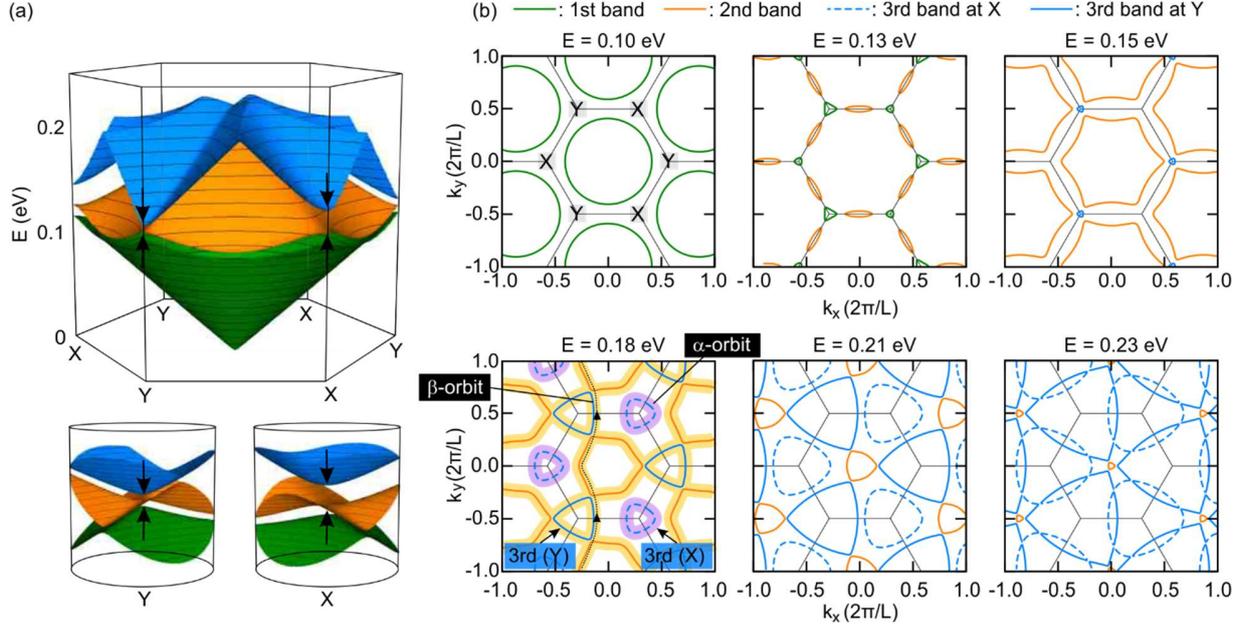

Figure 3. (a) Top: Three-dimensional plot of the first three conduction bands centered at the *K*-symmetry point of the reduced Brillouin zone (Fig. S3) calculated with the effective continuum model. Bottom: Band dispersion of the three bands in the top panel near the X and Y corners (Eq. S4, SI). (b) Energy contours of the first three conduction bands at six different Fermi energies. The green, orange, and blue lines show the contours of the first, second, and third bands, respectively. The pockets of the third band centered around X (blue dashed lines) are bounded by $\alpha$-orbits (purple shaded lines in the panel with $E = 0.18$ eV). These orbits are well separated in reciprocal space from the other isoenergetic contours and almost do not interact with them. The third miniband also has a second set of electron pockets centered around Y (blue solid lines). The bounding orbit in reciprocal space hybridizes with the Fermi surface of the second miniband in the presence of a magnetic field via magnetic breakdown (yellow shaded lines in the panel with $E = 0.18$ eV). This produces a hybrid, snake-like orbit with a much larger area in reciprocal space ($\beta$-orbit). One such a vertical snake orbit is marked with a dotted line and two arrows along its path.

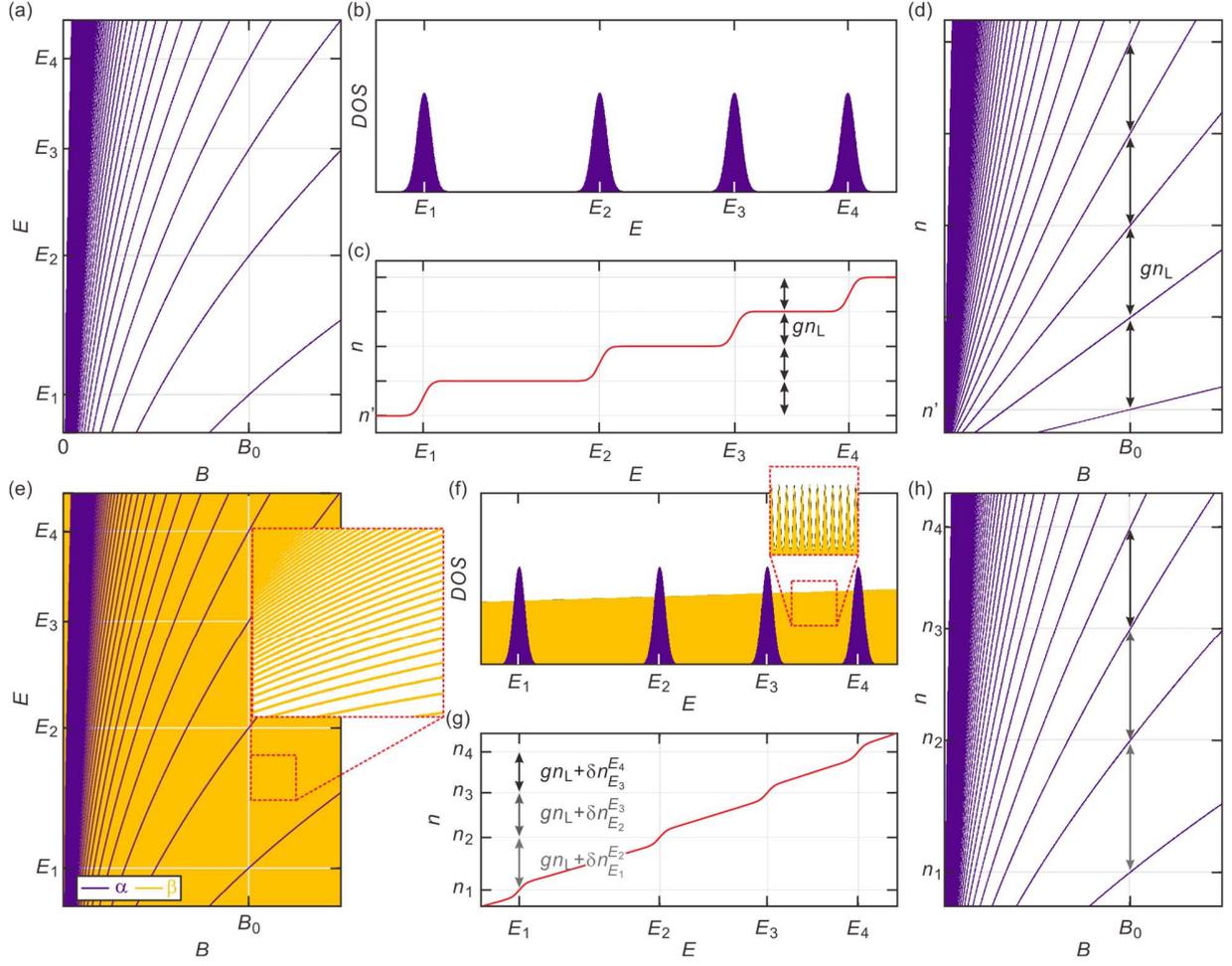

Figure 4. (a)-(d) Schematic energy spectrum and Wannier diagram of a system with a single set of Landau levels $\alpha$ (e.g., graphene monolayer). (a) Landau levels in the $E$-$B$ space. $E_i$ (i = 1, 2, 3, 4) represents the energies of the Landau levels at a field $B_0$. (b) Density of states and (c) electron density $n$ of the series of Landau levels in (a) at a field $B_0$ plotted against $E$, (d) Wannier diagram in the $n$-$B$ space. Note that regardless of the energy spacing between the levels in $\alpha$, $E_i$, in (b), the changes of electron densities $gn_L$, $n_L = eB/h$ is the electron density of each Landau level, and $g$ is the degeneracy of the bands, at all the gaps in (c) and (d) are the same. (e)-(h) Plots similar to (a)-(d) but for a system with two sets of Landau levels $\alpha$ and $\beta$ with very different energy spacing. Note that the electron density of $\beta$ between $E_i^\alpha$ and $E_{i+1}^\alpha$ ($\delta n_{E_i}^{E_{i+1}} = \int_{E_i^\alpha}^{E_{i+1}^\alpha} D_\beta(E) dE$) scales with the energy spacing between $E_i^\alpha$ and $E_{i+1}^\alpha$ (g), thus makes the energy spacing be embedded in the spacing in the Wannier diagram (h).

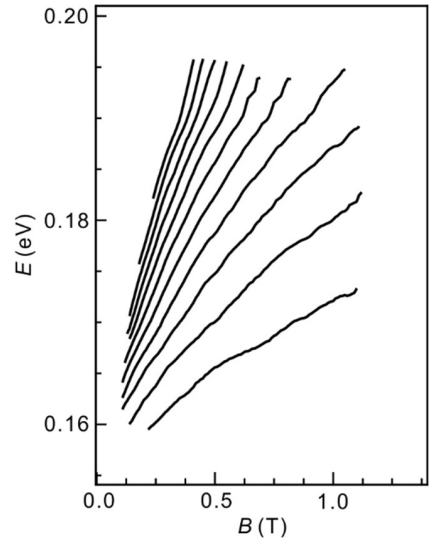

Figure 5. Energy spectrum of $\alpha$-orbits decoded by using Eq. (2) (see text) from the peaks of $\sigma_{xx}$ measured in experiment.

## Methods

*Device fabrication*

Device D1 is composed of a graphene monolayer encapsulated on either side with a hBN layer. The doped Si substrate covered with a thermal oxide serves as the backgate. The heterostructure was assembled using a viscoelastic stamp method that has been described in detail before [19]. We selected hBN flakes and a graphene flake that either had one very straight edge or two boundaries that form a 120° angle. For the top hBN, such a boundary was then intentionally aligned with the straight boundary of the graphene flake a motorized x,y,z and $\theta$ stage. A second moiré potential from the bottom hBN flake was avoided by intentionally placing the bottom hBN with respect to the graphene flake at a twist angle of at least 5°. For the seond device D2, the dry pick-up and transfer process using a PPC/PDMS stamp as described in Ref. [20] was used to assemble a hBN/graphite/hBN/graphene/hBN/graphite heterostructure. Device D2 was fabricated in vacuum ($5 \times 10^{-4}$ mbar) to increase the useable area of the layer stack [21, 22]. In order to reduce bubbles and wrinkles both devices were annealed for 30 min. at 500º C in forming gas at a pressure of 150 mbar. For further details of the fabrication procedure, we refer to Ref. [17, 19].

*Magnetotransport Measurements*

Magnetotransport data on D1 was recorded in a top-loading-into-mixture dilution refrigerator from Oxford Instruments at a base temperature of approximately 30 mK. D2 was measured in a dry Physical Property Measurement System from Quantum Design (PPMS Dynacool) for temperatures down to 1.7 K. The longitudinal and transverse resistance were measured in four terminal configuration using lock-in technique with an alternating current $I$ of 10 nA and a frequency $f$ of 17.777 Hz.

## ASSOCIATED CONTENT

**Supporting Information**

Extended magnetotransport data, estimate of the superlattice period and twist angle for device D1, magnetotransport on device D2, superlattice Brillouin zone, extended energy spectrum and Wannier diagram, effective Hamiltonian

## AUTHOR INFORMATION

**Corresponding Author**

**Pilkyung Moon** - Arts and Sciences, NYU Shanghai, Shanghai 200124, China and NYU-ECNU Institute of Physics at NYU Shanghai, Shanghai 200062, China; Email: pilkyung.moon@nyu.edu

**Mikito Koshino -** Department of Physics, Osaka University, Toyonaka 560-0043, Japan; Email: koshino@phys.sci.osaka-u.ac.jp,

**Jurgen. H. Smet** - Max-Planck-Institut für Festköperforschung, 70569 Stuttgart, Germany; Email: j.smet@fkf.mpg.de

**Authors**

**Youngwook Kim** - Max-Planck-Institut für Festköperforschung, 70569 Stuttgart, Germany; Department of Physics and Chemistry, DGIST, 42988 Daegu, Korea

**Kenji Watanabe** - Research Center for Functional Materials, National Institute for Materials Science, Tsukuba 305-0044, Japan

**Takashi Taniguchi** - International Center for Materials Nanoarchitectonics, National Institute for Materials Science, Tsukuba 305-0044, Japan

**Notes**

The authors declare no competing financial interest.

**Author Contributions**

P. M., Y. K., M. K. contributed equally

## ACKNOWLEDGEMENTS

We thank K. von Klitzing for fruitful discussions and Y. Stuhlhofer, S. Göres, and M. Hagel for assistance with sample preparation. P.M. acknowledges the support by National Science Foundation of China (Grant No. 12074260), NYU-ECNU Institute of Physics at NYU Shanghai, and NYU Shanghai (Super Boost Funds). The computation was carried out on the High Performance Computing resources at NYU Shanghai. Y. K acknowledges financial support from the Alexander von Humboldt Foundation. This work has been supported by the the DFG Priority Program SPP 2244 and the graphene flagship core 3 program. The work at DGIST is supported by the Basic Science Research Program NRF-2020R1C1C1006914, NRF-2022M3H3A1098408 through the National Research Foundation of Korea (NRF), the DGIST R&D program (23-CoE-NT-01) of the Korean Ministry of Science and ICT, and the BrainLink program funded by the Ministry of Science and ICT through the National Research Foundation of Korea (2022H1D3A3A01077468). We also acknowledge the partner group program of the Max Planck Society. K.W. and T.T. acknowledge support from the JSPS KAKENHI (Grant Numbers 20H00354, 21H05233 and 23H02052) and World Premier International Research Center Initiative (WPI), MEXT, Japan.)

# Supporting Information for "Non-linear Landau fan diagram for graphene electrons exposed to a Moiré potential"

Pilkyung Moon[1,2, △,*], Youngwook Kim[3,4, △], Mikito Koshino[5,△,*], Takashi Taniguchi[6], Kenji Watanabe[7], and Jurgen H. Smet[3,*]

[1]*Arts and Sciences, NYU Shanghai, Shanghai 200124, China*

[2]*NYU-ECNU Institute of Physics at NYU Shanghai, Shanghai 200062, China*

[3]*Max-Planck-Institut für Festköperforschung, 70569 Stuttgart, Germany*

[4]*Departement of Physics and Chemistry, DGIST, 42988, Korea*

[5]*Department of Physics, Osaka University, Toyonaka 560-0043, Japan*

[6]*International Center for Materials Nanoarchitectonics, National Institute for Materials Science, Tsukuba 305-0044, Japan*

[7]*Research Center for Functional Materials, National Institute for Materials Science, Tsukuba 305-0044, Japan*

△These authors contributed equally

*E-mail: pilkyung.moon@nyu.edu, koshino@phys.sci.osaka-u.ac.jp, and j.smet@fkf.mpg.de

## S1. EXTENDED MAGNETOTRANSPORT DATA

The color legends and the maximum magnetic field in the top panels of Fig. 1a and 1b were selected so as to enhance the visibility of the non-linear trajectories traced by the conductivity minima at high electron densities ($V_g > 50\ V$), that are central to this work. The complete data set up to 21.5 T recorded on device D1 is shown in Fig. S1 using an appropriately adjusted color legend. The conventional linear Landau fans that emerge from the main and mini charge neutrality points at zero field are better visible in this extended data set. Integer quantum Hall states are observed from 0.5 T and broken symmetry states appear near 3 T. The commensurability between the moiré superlattice unit cell and the magnetic unit cell produces features associated with the fractal Hofstadter energy spectrum. For instance, the horizontal white lines in $\sigma_{xx}$ signal high conductivity due to the formation of energy bands with a finite width at commensurate fields, where the normalized magnetic flux $\phi/\phi_0$ ($\phi=BA$, $\phi_0 = h/e$) takes on a rational number $p/q$. Here, $p$ and $q$ are coprime integers. More than 80 quantum Hall states associated with the fractal spectrum can be identified starting from 2 T in this color map. All of these features are flux linear.

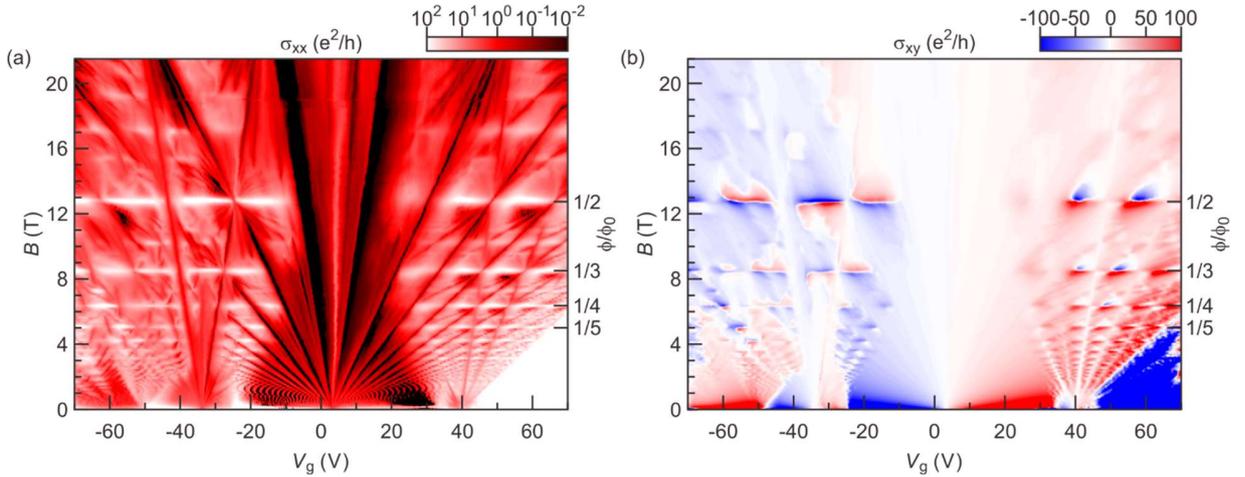

Figure S1. Extended data set for device D1 up to 21.5 T complementing Fig. 1 in the main text. The color legend was adjusted and optimized for improved visibility of the linear Landau fans and the Hofstadter butterfly features. The dimensionless magnetic field, $\phi/\phi_0$, is used as the right ordinate due to its relevance for the fractal butterfly spectrum. Here, $\phi$ is the flux contained in the moiré superlattice unit cell and $\phi_0$ is the magnetic flux quantum, ($h/e$).

## S2. ESTIMATE OF THE SUPERLATTICE PERIOD AND TWIST ANGLE FOR DEVICE D1

The size of the moiré superlattice unit cell or twist angle can be estimated from the density required to go from the main CNP at the Dirac point to the mini CNP when the first miniband is completely filled. For device D1 this requires a change in the gate voltage of approximately 36.2 V. The density at this gate voltage is obtained from an analysis of the low magnetic field Shubnikov de-Haas oscillation frequency or alternatively the filling factor of quantum Hall states at a high magnetic field. It is approximately equal to $2.49 \times 10^{12}$ cm$^{-2}$ and should correspond to $g_s g_v n_0$, where $n_0$ is the inverse superlattice unit cell area = $[3^{-1/2}/2 \times a^2]^{-1}$. Here, $a$ denotes the lattice constant of the superlattice unit cell and $g_s g_v$ (=4) refers to the internal degrees of freedom due to spin and valley.

The expression yields a superlattice unit cell of approximately 13.9 nm for device D1 or a twist angle between hBN and graphene of 0°±0.1. The Hofstadter butterfly features also allow to identify at what magnetic field a single magnetic flux quantum penetrates the superlattice unit cell. This occurs at a field of approximately 26 T, close to the expected 25.2 T for a zero twist angle graphene/hBN heterostructure.

A second method to determine the superlattice periodicity is the use of Brown-Zak oscillations in Figs. S1a and S1b that appear at the simplest rational combinations $p/q$ of the normalized flux $\phi/\phi_0$, namely $1/n$ where $n$ is an integer. These quantum oscillation are density independent, since the total flux $\phi = B \times [3^{-1/2}/2 \times a^2]^{-1}$ and just depends on the size of the superlatiice unit cell. A best fit to the data yields $a$ = 13.7 nm. This value corresponds to a twist angle of 0.15°±0.05.

There is another alternative way to estimate the size of the superlattice unit cell using the Diophantine relation. It is described in Section S3 below.

## S3. MAGNETOTRANSPORT ON DEVICE D2

Device D2 consists of a hBN/graphene/hBN heterostructure sandwiched between both a top and bottom graphite gate. These additional graphite layers have been reported as beneficial for the overall sample quality [S1]. They reduce disorder fluctuations thereby allowing the observation of interaction induced states as well as conventional quantum Hall states at relatively low magnetic fields.

Fig. S2a shows the dependence of the longitudinal resistance on the back-gate voltage at $T = 1.7$ K and $B = 0$ T. The resistance exhibits a sharp maximum at 0 V corresponding to the main CNP as well as two satellite peaks near $\pm 6$ V when the lowest miniband is completely filled or emptied. A very rich sequence of quantum Hall states due to the Hofstdter butterfly spectrum can be clearly seen. The field and density coordinates, where these states appear, obey the Diophantine equation: $(n/n_0) = t \times (\phi/\phi_0) + s$. Here, $t$ is the Chern number and $s$ represents the index for Bloch band filling. The periodicity of the moiré superlattice period can be obtained from the crossing of the fan line with $t = -6$ from the mini CNP on the electron side and the fan line with $t = 6$ from the charge neutrality point. These two quantum Hall states cross at $B = 12$ T as shown in Figure S2b. From the Diophantine relation, we can construct the following relation, $(n/n_0) = 6(\phi/\phi_0) = -6(\phi/\phi_0) + 4$, hence $\phi/\phi_0 = 1/3 = B \times$ (unit cell area) $\times e/h$. This implies that the superlattice periodicity, $a$, equals 12.3 nm. This corresponds to a twist angle of 0.6° between hBN and graphene.

Figure S2c shows a data set recorded for a smaller range of magnetic fields between -3 T and +3 T in the regime of large electron densities. In the vicinity of $V_g = 6$ V, multiple quantum Hall states appear near the mini-CNP. At even higher densities, the resistance shows multiple minima that follow non-linear trajectories. Two of these are highlighted with white solid lines. To emphasize the non-linearity, white dashed lines are included that are linear extrapolations of these two resistance minima from high field down to zero field. The non-linear features behave similar as in device D1. Their strength weakens with increasing magnetic field and finally disappear near 3 T.

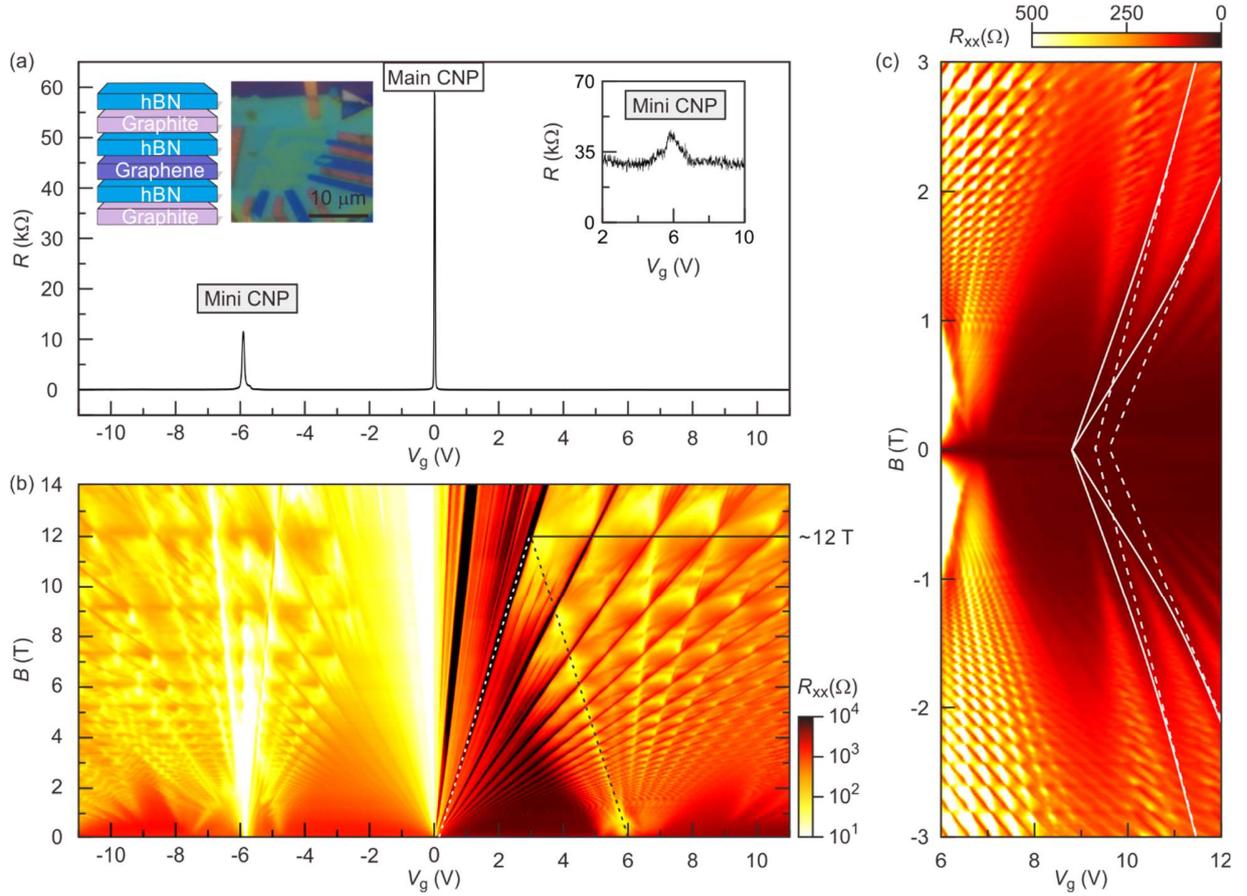

**Figure S2**. Magnetoresisitance data recorded on device D2. (a) Longitudinal resistance $R_{xx}$ as a function of back-gate voltage for B = 0T. The resistance peaks when the highest valence miniband is completely emptied of electrons, the lowest conduction miniband is completely occupied with electrons or when the chemical potential crosses the Dirac point. Left inset: Schematic of the device structure as well as an optical image. The scale bar equals 10 μm. Right inset: Enlargement of the weak resistance peak when the lowest conduction miniband gets fully occupied. (b) Longitudinal magnetoresistance as a function of field and density for fields up to 14 T. The data set is dominated by Hofstadter butterfly features. The white dashed line marks the quantum Hall feature for $t = 6$ in the Diophantine equation. It emanates from the main charge neutrality point. The black dashed line corresponds to the quantum Hall feature for $t = -6$. It emerges from the mini CNP of the lowest conduction miniband instead. These lines cross near 12 T (black horizontal line). (c) Magnetoresistance as a function back-gate voltage $V_g$ and field $B$ for a smaller range of magnetic fields and the regime of high electron density. The field is swept from −3 T to +3 T in 10 mT step, whereas the back-gate voltage is varied from 6 V to 12 V in 1 mV steps. Two of the resistance minima that follow a non-linear trajectory are highlighted with solid white lines. The dashed lines are linear extrapolations of the position of these resistance minima at high fields.

## S4. SUPERLATTICE BRILLOUIN ZONE

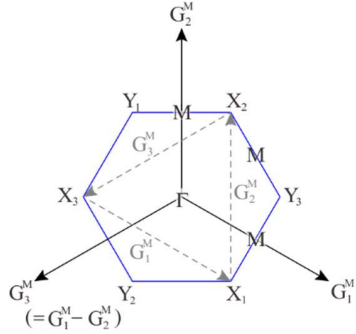

**Figure S3**. Superlattice Brillouin zone of the graphene/hBN moiré superlattice spanned by the reciprocal lattice vectors $\mathbf{G}_i^M$. $X_i$ and $Y_i$ ($i$ = 1, 2, 3) denote the two inequivalent corners of the superlattice Brillouin zone, aka mini CNP point. Note that $\mathbf{G}_i^M$ connects the three equivalent corners $X_i$ (and $Y_i$ as well). Thus, the moiré superlattice potential couples the Dirac spinors at these corners.

## S5. EXTENDED ENERGY SPECTRUM AND WANNIER DIAGRAM

Figures S4a and S4b show the plots similar to Figs. 2c and 2d but for a wider range of electron energy and density. Both figures clearly show the linear trajectories of the minima in the density of states that converge to the band edges at $n/n_0$ = 0, 4 as well as the non-linear trajectories starting from $n/n_0 > 5$.

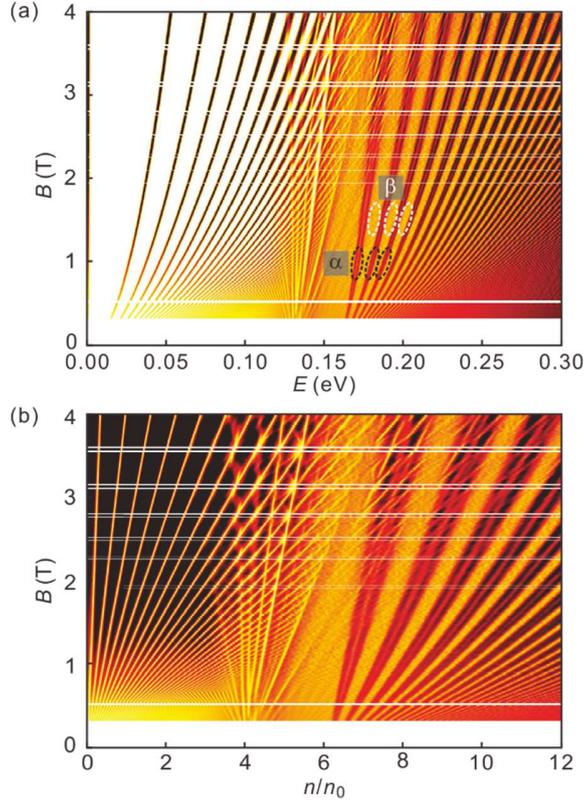

**Figure S4**. Plot of the density of states across of the ($B$,$E$) and ($B$,$n/n_0$) parameter space as in Figs. 2c and 2d of the main text, but for a wide range of electron energies and densities.

## S6. EFFECTIVE HAMILTONIAN

In the main text and Fig. 4, we attribute the presence of two different series of Landau levels associated with the $\alpha$- and $\beta$-orbits to the strong asymmetry in the energy dispersions near the $X$ and $Y$ corners of the reduced Brillouin zone. The isoenergetic contours of the third conduction miniband near the $X$ corners are well separated in reciprocal space from the isoenergetic contours of the other minibands. This gives rise to a discrete set of Landau levels. However, isoenergetic contours of the third conduction miniband near the $Y$ corner nearly touch the isoenergetic contours of the second conduction miniband, giving rise to an almost continuous energy spectrum. This distinction between the contours near $X$ and $Y$ originates from the different coupling strength at these corners. The Hamiltonian around the zone corners $Z_i^\eta$ ($i = 1,2,3$) ($Z_i^\eta$ is $X_i$ for $\eta = 1$ and $Y_i$ for $\eta = -1$) near the $K$-valley of monolayer graphene can be written as

$$H = IE_0^Z + \begin{pmatrix} s\hbar v\tilde{k}\cos\left(\theta + \frac{\pi}{3}\right) & \alpha^* + \beta^*\tilde{k}\cos\theta & \alpha + \beta\tilde{k}\cos\left(\theta + \frac{2\pi}{3}\right) \\ \alpha + \beta\tilde{k}\cos\theta & s\hbar v\tilde{k}\cos\left(\theta - \frac{\pi}{3}\right) & \alpha^* + \beta^* k\cos\left(\theta - \frac{2\pi}{3}\right) \\ \alpha^* + \beta^*\tilde{k}\cos\left(\theta + \frac{2\pi}{3}\right) & \alpha + \beta k\cos\left(\theta - \frac{2\pi}{3}\right) & -s\hbar v\tilde{k}\cos\theta \end{pmatrix} \quad (S1)$$

in the basis of $|Z_1^\eta + \tilde{k}\rangle, |Z_2^\eta + \tilde{k}\rangle, |Z_3^\eta + \tilde{k}\rangle$ and up to the first order in $\tilde{k}$. Here, $|k\rangle$ is the wave function of unproximitized graphene for the vector $k$, where $\tilde{k} = \tilde{k}(\cos\theta, \sin\theta)$ is the wave vector measured from each $Z_i^\eta$, $E_0^Z = s\hbar vG^M/\sqrt{3}$ is the energy of unproximitized graphene at $Z_i^\eta$, $s$ is 1 for the conduction band and -1 for the valence band, $v$ is the Fermi velocity of monolayer graphene, $G^M = |G_i^M|$, $I$ is the $3 \times 3$ identity matrix, and

$$(\alpha, \beta) = \begin{cases} V_1 e^{i\eta\psi}(\omega^*/2, (-\omega + 2\omega^*)\gamma) & (s = 1, \eta = 1) \\ (0, 0) & (s = 1, \eta = -1) \\ V_1 e^{i\eta\psi}(-3\omega^*/2, 3\omega\gamma) & (s = -1, \eta = 1) \\ V_1 e^{i\eta\psi}(2, 2(1 - 2\omega^*)\gamma) & (s = -1, \eta = -1) \end{cases}, \quad (S2)$$

where $V_1 \approx 0.0210$ eV, $\psi \approx -0.29$ (rad), and $\gamma = \sqrt{3}/2G^M$. A similarity transformation of $H$ with

$$U = \frac{1}{\sqrt{3}}\begin{pmatrix} 1 & 1 & 1 \\ 1 & \omega & \omega^* \\ 1 & \omega^* & \omega \end{pmatrix} \quad (S3)$$

gives

$H' = U^\dagger HU = IE_0^Z +$

$$\begin{pmatrix} \alpha + \alpha^* & -\frac{\omega^*}{2}(s\hbar v - \beta^*\omega^* - \beta\omega)\tilde{k}e^{i\theta} & -\frac{\omega}{2}(s\hbar v - \beta\omega^* - \beta^*\omega)\tilde{k}e^{-i\theta} \\ -\frac{\omega}{2}(s\hbar v - \beta\omega - \beta^*\omega^*)\tilde{k}e^{-i\theta} & \alpha\omega^* + \alpha^*\omega & -\frac{\omega^*}{2}(s\hbar v - \beta^* - \beta)\tilde{k}e^{i\theta} \\ -\frac{\omega^*}{2}(s\hbar v - \beta^*\omega - \beta\omega^*)\tilde{k}e^{i\theta} & -\frac{\omega}{2}(s\hbar v - \beta - \beta^*)\tilde{k}e^{-i\theta} & \alpha\omega + \alpha^*\omega^* \end{pmatrix}$$

$$(S4)$$
.

The bottom panels in Fig. 3a show the energy dispersion of the conduction bands ($s = 1$) plotted against $\tilde{k}$ near $X$ ($\eta = 1$) and $Y$ ($\eta = -1$), respectively. At the zone corner ($\tilde{k} = 0$), the eigenvalues are $E = \{E_0^Z + \alpha + \alpha^*, E_0^Z + \alpha\omega^* + \alpha^*\omega, E_0^Z + \alpha\omega + \alpha^*\omega^*\} = \{E_0^Z +$ $CV_1 \cos\psi$, $E_0^Z + CV_1 \cos(\psi + 2\pi/3)$, $E_0^Z + CV_1 \cos(\psi - 2\pi/3)$}, where $C$ is 1 ($-3$) and 0 (4) for $X$ and $Y$ in the conduction (valence) bands, respectively. Thus, similar to the numerical calculation of the band structures (top panel in Fig. 3a), the first three conduction bands at $Y$ are degenerate, while the bands at $X$ ({$E_0^Z - 0.015, E_0^Z - 0.005, E_0^Z + 0.220$}), especially the third band, are split by a finite amount. Thus, the electron pocket of the third band at $Y$ tunnels to the second band and is hybridized to a band with very large area $S$, while the pocket encircling $X$ works as an almost independent band with small $S$. To be more precise, the interaction that is of higher order in $\mathbf{G}_i^M$ opens a tiny pseudogap at this point, but this small opening does not prevent the band hybridization discussed in the main text.

The dispersion of the valence minibands ($s = -1$) in Fig. S5 shows that these bands are well separated in energy by the large band gap at the mini CNP points, unlike what happens to the conduction minibands (Fig. 3a).

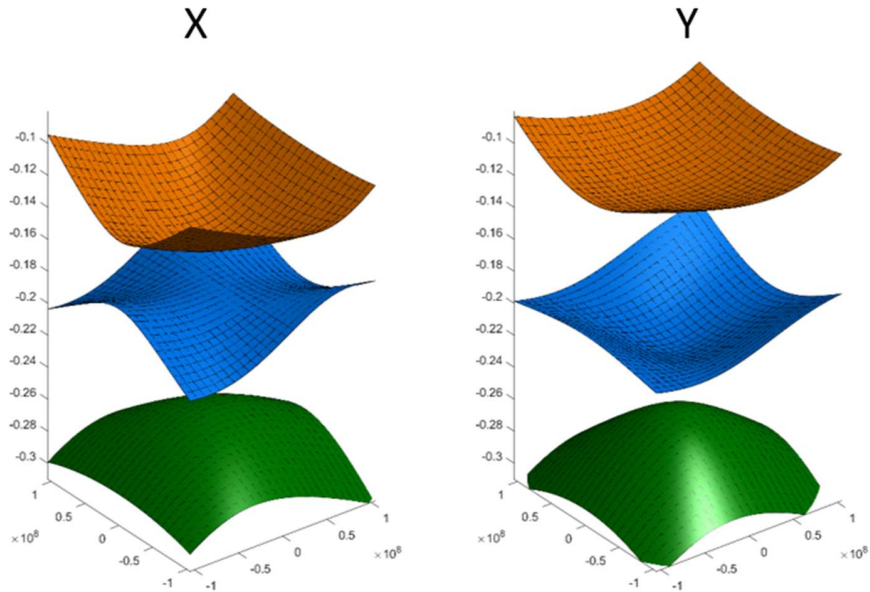

**Figure S5**. Dispersion near the $X$ (left) and $Y$ (right) corners of the three topmost valence minibands, calculated by the effective model (Eq. S4). The minibands are well separated and do not overlap.

**Reference for Supporting Information**